\documentclass[aps,prd,reprint,footnoteinbib,longbibliography]{revtex4-2}
\pdfoutput=1
\usepackage[T1]{fontenc}
\usepackage{bm,amsmath,amssymb,color,bbm,mathrsfs} 
\usepackage[pdftex]{graphicx}
\usepackage[colorlinks=true, pdfstartview=FitH, linkcolor=blue, citecolor=blue, urlcolor=blue]{hyperref}

\DeclareMathOperator\Tr{Tr}
\DeclareMathOperator\Pf{Pf}
\DeclareMathOperator\rme{\mathrm{e}}

\renewcommand{\bar}[1]{\overline{#1}}

\newcommand{\bep}{\begin{pmatrix}} 
\newcommand{\eep}{\end{pmatrix}}

\newcommand{\SU}{\text{SU}}
\newcommand{\SO}{\text{SO}}
\renewcommand{\O}{\text{O}}
\newcommand{\U}{\text{U}}
\newcommand{\1}{\mathbbm{1}}
\newcommand{\RR}{\mathbb{R}}
\newcommand{\CC}{\mathbb{C}}

\renewcommand{\epsilon}{\varepsilon}
\newcommand{\rmd}{\mathrm{d}}

\newcommand{\m}{\hat{m}}
\newcommand{\n}{\mathbf{n}}

\def\ba#1\ea{\begin{align}#1\end{align}}
\def\mkakko#1{\left( #1 \right)}
\def\ckakko#1{\left\{ #1 \right\}}
\def\kkakko#1{\left[ #1 \right]}

\begin{document}
\title{Chiral random matrix theory for single-flavor spin-one Cooper pairing}

\author{Takuya Kanazawa}
\affiliation{Research and Development Group, Hitachi, Ltd., Kokubunji, Tokyo 185-8601, Japan}
\allowdisplaybreaks

\begin{abstract}
We propose a new non-Hermitian chiral random matrix model that describes single-flavor spin-one Cooper pairing of quarks. For three colors the model shows spontaneous breaking of color $\SU(3)_C$ and spin $\SO(3)_J$ symmetries down to the diagonal $\SO(3)_{C+J}$ subgroup, in striking analogy to the color-spin locked phase of one-flavor QCD at high density. For two colors, color-singlet spin-one diquarks condense and trigger symmetry breaking $\U(1)\times\SO(3)_J\to\SO(2)_J$. In both cases the microscopic large-$N$ limit is rigorously taken and the effective theory of Nambu-Goldstone modes is derived. 
\end{abstract}
\maketitle

\section{Introduction}

In relativistic cold ultradense matter, the Fermi surface is destabilized by attractive interactions between quarks and the ground state is likely to exhibit color superconductivity \cite{Bailin:1983bm}. For three flavors, the most stable state at asymptotically high density is believed to be the so-called color-flavor locked (CFL) phase in which color and flavor symmetries are all spontaneously broken \cite{Alford:1998mk}, while for two flavors the 2SC phase breaks color down to $\SU(2)$ without chiral symmetry breaking \cite{Bailin:1983bm,Alford:1997zt,Rapp:1997zu}. In reality, however, constraints from color and electric charge neutrality and $\beta$-equilibrium as well as the large strange quark mass enhance the mismatch of Fermi momenta of different flavors, and make the CFL and 2SC phases less favored at moderate densities that can be reached in the interior of compact stars. Various alternative phases are proposed in the literature \cite{Rajagopal:2000wf,Alford:2001dt,Buballa:2003qv,Shovkovy:2004me,Alford:2007xm,Fukushima:2010bq,Fukushima:2013rx,Casalbuoni:2018haw}. The possibility of Cooper pairing between equal flavors is particularly interesting because it is free from stress under the mismatch of Fermi momenta. Due to the Pauli principle, the Cooper pair wave function must be antisymmetric under the exchange of quantum numbers of quarks. The perturbative one-gluon exchange provides an attractive interaction in the color antisymmetric channel, so it seems reasonable to assume that the Cooper pair is antisymmetric in colors. Then Lorentz-scalar pairing is forbidden and the Cooper pair is forced to carry a nonzero total angular momentum $J$. Studies by various authors \cite{Bailin:1983bm,Iwasaki:1994ij,Schafer:1999pb,Schafer:2000tw,Alford:2002rz,Schmitt:2003xq,Schmitt:2004hg,Schmitt:2004et,Schmitt:2005ee,Aguilera:2005tg,Marhauser:2006hy,Feng:2007bg,Feng:2008dh,Brauner:2008ma,Feng:2009vt,Pang:2010wk} suggest that the ground state for $J=1$ at asymptotically high density is the color-spin locked (CSL) phase \cite{Bailin:1983bm,Schafer:2000tw}, where the color $\SU(3)_C$ and the spin $\SO(3)_J$ symmetries are dynamically broken to the diagonal $\SO(3)_{C+J}$ subgroup. This phase retains invariance under a combination of color and spatial rotation.  

Random matrix theory (RMT) \cite{Mehtabook,AGZbook,Akemann:2011csh} offers a powerful approach to probing nonperturbative dynamics of gauge theories. It is well established that statistical fluctuations of the Dirac operator eigenvalues in the microscopic domain are exactly described by RMT with chiral symmetry \cite{Verbaarschot:2000dy,Verbaarschot:2005rj}; the Dirac eigenvalues of order $1/V_4$ with $V_4$ the Euclidean spacetime volume correspond to the hard-edge limit of the Wishart-Laguerre ensemble. This correspondence generalizes to a weakly non-Hermitian Dirac operator at small chemical potential $\mu$ of order $1/\sqrt{V_4}$ \cite{Osborn:2004rf,Akemann:2004dr,Akemann:2007rf}. Moreover, in QCD-like theories (i.e., two-color QCD, QCD with adjoint quarks, and QCD with isospin chemical potential), there exist strongly non-Hermitian random matrix ensembles that govern the Dirac spectrum at high density $\mu\gg m_{\pi}$ \cite{Kanazawa:2009en,Akemann:2010tv,Kanazawabook,Kanazawa:2014lga}. The typical scale of the microscopic domain in these theories is set by the BCS gap of quarks \cite{Kanazawa:2012zr}. The question whether universality of the microscopic limit of RMT holds for color-superconducting phases of dense three-color QCD has remained open \footnote{Phenomenological applications of RMT to color-superconducting phases of QCD can be found in \cite{Vanderheyden:1999xp,Vanderheyden:2000ti,Pepin:2000pv,Vanderheyden:2011iq,Sano:2011xs}. These works employed RMT only to study the general topology of the QCD phase diagram and did not discuss the microscopic large-$N$ limit.}. Recently a suitable matrix model was constructed for two and three flavors \cite{Kanazawa:2020dju}. 

In this paper, we propose a new non-Hermitian RMT that describes single-flavor Cooper pairing for two and three colors. By taking the microscopic large-$N$ limit with $N$ the matrix size we show, for three colors, that diquark condensation occurs in the color-triplet spin-one channel and spontaneously breaks $\U(1)_B\times\U(1)_A\times\SU(3)_C\times\SO(3)_J$ symmetry down to $\SO(3)_{C+J}$, which is the diagonal subgroup of $\SU(3)_C\times\SO(3)_J$. This is the first-ever realization of color-spin locking in RMT. For two colors, we show that color-singlet spin-one diquarks condense and break $\U(1)_B\times\U(1)_A\times\SO(3)_J$ spontaneously. There seems to be no previous work on one-flavor two-color QCD at high density and this work provides the first systematic study based on symmetries. 

This paper is organized as follows. In section~\ref{sc:mmdef} our new matrix model is defined and basic properties are summarized. In section~\ref{sc:Nc3} the large-$N$ limit is taken for three colors and color-spin locking is demonstrated. Similarities and differences between RMT and QCD are clarified. In section~\ref{sc:Nc2} the analysis for two colors is presented. The breaking of rotational symmetry by diquark condensates is demonstrated and the effective theory of soft modes is rigorously derived. We conclude in section~\ref{sc:conc}.

\section{\label{sc:mmdef}The matrix model}

The partition function of our random matrix model for $N_c$ colors is defined by
\ba
	Z & = \int \rmd X \int \rmd Y \int \prod_{A=0}^{N_c^2-1}\rmd V^A \int \prod_{A=0}^{N_c^2-1}\rmd W^A 
	\notag
	\\
	& \quad \times 
	\exp\big\{-N\Tr(X^T X) - N\Tr(Y^TY) 
	\notag
	\\
	& \quad 
	-2N \Tr[(V^A)^TV^A]-2N \Tr[(W^A)^TW^A]\big\}
	\notag
	\\
	& \quad \times \det(\mathscr{D}+m\1_{4N_cN}) 
\ea
where $X,Y,V^A,W^A(A=0,1,\cdots,N_c^2-1)$ are $N\times N$ real matrices and $\rmd X,\rmd Y, \rmd V^A, \rmd W^A$ are flat Cartesian measures over $\RR^{N\times N}$. The ``Dirac operator'' $\mathscr{D}$ is a $4N_cN\times 4N_cN$ matrix defined as
\ba
	&\qquad \mathscr{D}=\begin{pmatrix}0&D_R\\D_L&0\end{pmatrix}
	\\
	D_L & \equiv (V^A\otimes t^A + i X\otimes \1_{N_c}) \otimes \1_2 
	\label{eq:D_L}
	\\
	D_R & \equiv (W^A\otimes t^A + i Y\otimes \1_{N_c}) \otimes \1_2
	\label{eq:D_R}
\ea
where $t^A(A=0,1,\cdots,N_c^2-1)$ are the generators of $\U(N_c)$ in the fundamental representation, normalized as $\Tr(t^At^B)=2\delta_{AB}$. The last factor $\1_2$ in \eqref{eq:D_L} and \eqref{eq:D_R} indicates that $D_{L,R}$ are diagonal in the spin space. In the chiral limit, there is a symmetry
\ba
	[\SU(N_c)\times\SU(2)_J]_L \times [\SU(N_c)\times\SU(2)_J]_R
\ea
which is reduced to the vectorial subgroup $[\SU(N_c)\times\SU(2)_J]_V$ by nonzero quark mass. There is also an orthogonal symmetry
\ba
	\begin{split}
	&V^A\to g_1V^A g_2^T, \quad X \to g_1X g_2^T,
	\\
	&W^A\to g_2W^A g_1^T, \quad Y \to g_2 Y g_1^T
	\end{split}
\ea
with $g_{1,2}\in\O(N)$.

Let us introduce quarks $\psi^\alpha_{ai}$ and antiquarks $\bar\psi^\alpha_{ai}$, where $a\in\{1,\cdots,N_c\}$ is color, $i\in\{1,2\}$ is spin and $\alpha\in\{1,2,\cdots,N\}$. Then
\begin{widetext}
\ba
	Z & = \int \rmd \bar\psi_R \rmd \bar\psi_L \rmd \psi_R \rmd \psi_L 
	\int \rmd X \int \rmd Y \int \prod_{A=0}^{N_c^2-1}\rmd V^A \int \prod_{A=0}^{N_c^2-1}\rmd W^A 
	\notag
	\\
	& \quad \times 
	\exp\ckakko{-N\Tr(X^T X) - N\Tr(Y^TY) -2N \Tr[(V^A)^TV^A]-2N \Tr[(W^A)^TW^A]}
	\notag
	\\
	& \quad \times 
	\exp\kkakko{
		\begin{pmatrix}\bar\psi_R \\ \bar\psi_L\end{pmatrix}^\alpha_{ai}
		\begin{pmatrix} m\delta_{\alpha\beta}\delta_{ab} & 
		W^A_{\alpha\beta} t^A_{ab} + i Y_{\alpha\beta}\delta_{ab} \\
		V^A_{\alpha\beta} t^A_{ab} + i X_{\alpha\beta} \delta_{ab} & 
		m^*\delta_{\alpha\beta}\delta_{ab} \end{pmatrix}
		\begin{pmatrix}\psi_L \\ \psi_R\end{pmatrix}^\beta_{bi}
	}
	\\
	& = \int \rmd \bar\psi_R \rmd \bar\psi_L \rmd \psi_R \rmd \psi_L 
	\int \rmd X \int \rmd Y \int \prod_{A=0}^{N_c^2-1}\rmd V^A \int \prod_{A=0}^{N_c^2-1}\rmd W^A 
	\notag
	\\
	& \quad \times \exp\big[
	-NX_{\alpha\beta}^2-NY_{\alpha\beta}^2-2N(V^A_{\alpha\beta})^2-2N(W^A_{\alpha\beta})^2
	+m\bar\psi_{Rai}^\alpha\psi_{Lai}^\alpha+m^*\bar\psi_{Lai}^\alpha\psi_{Rai}^\alpha
	\notag
	\\
	& \quad +
		\bar\psi^\alpha_{Lai}V^A_{\alpha\beta} t^A_{ab}\psi^\beta_{Lbi} 
		+ i \bar\psi^\alpha_{Lai}X_{\alpha\beta} \psi^\beta_{Lai} 
		+ \bar\psi^\alpha_{Rai}W^A_{\alpha\beta} t^A_{ab}\psi^\beta_{Rbi} 
		+ i \bar\psi^\alpha_{Rai}Y_{\alpha\beta}\psi^\beta_{Rai}
	\big]\,.
\ea
\end{widetext}
It is now straightforward to integrate out the Gaussian random matrices. Using the completeness relation $t^A_{ab}t^A_{cd}=2\delta_{ad}\delta_{bc}$ we obtain
\ba
	Z & \propto \int \rmd \bar\psi_R \rmd \bar\psi_L \rmd \psi_R \rmd \psi_L 
	\exp\Big[
		m\bar\psi_{Rai}^\alpha\psi_{Lai}^\alpha+m^*\bar\psi_{Lai}^\alpha\psi_{Rai}^\alpha 
	\notag
	\\
	& \quad 
		- \frac{1}{4N}\bar\psi^\alpha_{Lai}\bar\psi^\alpha_{Lbj}\psi^\beta_{Lbj}\psi^\beta_{Lai}
		+ \frac{1}{4N}\bar\psi^\alpha_{Lai}\bar\psi^\alpha_{Lbj}\psi^\beta_{Laj}\psi^\beta_{Lbi}
	\notag
	\\
	& \quad 
		+ (L\leftrightarrow R)
	\Big]
	\\
	& = \int \rmd \bar\psi_R \rmd \bar\psi_L \rmd \psi_R \rmd \psi_L 
	\exp\Big[
		m\bar\psi_{Rai}^\alpha\psi_{Lai}^\alpha+m^*\bar\psi_{Lai}^\alpha\psi_{Rai}^\alpha
	\notag
	\\
	& \quad 
		+ \frac{1}{4N}\bar\psi^\alpha_{Lai}\bar\psi^\alpha_{Lbj}\psi^\beta_{Lcj}\psi^\beta_{Ldi}
		(\delta_{ac}\delta_{bd}-\delta_{ad}\delta_{bc})
		+ (L\leftrightarrow R)
	\Big]
	\\
	& = \int \rmd \bar\psi_R \rmd \bar\psi_L \rmd \psi_R \rmd \psi_L 
	\exp\Big[
		m\bar\psi_{Rai}^\alpha\psi_{Lai}^\alpha+m^*\bar\psi_{Lai}^\alpha\psi_{Rai}^\alpha
	\notag
	\\
	& \quad 
		+ \frac{1}{8N}\bar\psi^\alpha_{Lai}\bar\psi^\alpha_{Lbj}\psi^\beta_{Lck}\psi^\beta_{Ld\ell}
		\cdot 2\delta_{jk}\delta_{i\ell} \cdot 
		(\delta_{ac}\delta_{bd}-\delta_{ad}\delta_{bc})
	\notag
	\\
	& \quad 
		+ (L\leftrightarrow R)
	\Big]\,.
\ea
For the Pauli matrices $\sigma_M\equiv(\1_2,\sigma_i)$ there is the identity
\ba
	\sum_{M=0}^{3}(\sigma_M\sigma_2)_{ij}(\sigma_2\sigma_M)_{k\ell} = 2\delta_{jk}\delta_{i\ell}\,.
\ea
It allows us to write
\ba
	Z & \propto 
	\int \rmd \bar\psi_R \rmd \bar\psi_L \rmd \psi_R \rmd \psi_L 
	\exp\Big[
		m\bar\psi_{Rai}^\alpha\psi_{Lai}^\alpha+m^*\bar\psi_{Lai}^\alpha\psi_{Rai}^\alpha
	\notag
	\\
	& \quad 
	+ \frac{1}{8N}(\bar\psi^\alpha_{La} \sigma_M\sigma_2 \bar\psi^\alpha_{Lb})
	(\psi^\beta_{Lc} \sigma_2\sigma_M \psi^\beta_{Ld})
	(\delta_{ac}\delta_{bd}-\delta_{ad}\delta_{bc})
	\notag
	\\
	& \quad + (L\leftrightarrow R)
	\Big]
\ea
where we have omitted the spinor indices for brevity. Noting that the $M=0$ term vanishes identically due to the Pauli principle, we are left with
\ba
	Z & \propto 
	\int \rmd \bar\psi_R \rmd \bar\psi_L \rmd \psi_R \rmd \psi_L 
	\exp\Big[
		m\bar\psi_{Rai}^\alpha\psi_{Lai}^\alpha+m^*\bar\psi_{Lai}^\alpha\psi_{Rai}^\alpha
	\notag
	\\
	& \quad 
	+ \frac{1}{8N}(\bar\psi^\alpha_{La} \sigma_m\sigma_2 \bar\psi^\alpha_{Lb})
	(\psi^\beta_{Lc} \sigma_2\sigma_m \psi^\beta_{Ld})
	(\delta_{ac}\delta_{bd}-\delta_{ad}\delta_{bc})
	\notag
	\\
	& \quad + (L\leftrightarrow R)
	\Big]
	\label{eq:Z87}
\ea
with $m\in\{1,2,3\}$, which should not be confused with the quark mass. The treatment henceforth depends on the number of colors and is discussed in the following sections.

\section{\label{sc:Nc3}Three colors}

For $N_c=3$ the identity $\epsilon_{abe}\epsilon_{cde}=\delta_{ac}\delta_{bd}-\delta_{ad}\delta_{bc}$ holds. Hence we have from \eqref{eq:Z87}
\ba
	Z & \propto 
	\int \rmd \bar\psi_R \rmd \bar\psi_L \rmd \psi_R \rmd \psi_L 
	\exp\Big[
		m\bar\psi_{Rai}^\alpha\psi_{Lai}^\alpha+m^*\bar\psi_{Lai}^\alpha\psi_{Rai}^\alpha
	\notag
	\\
	& \quad 
	+ \frac{1}{8N}(\bar\psi^\alpha_{La} \sigma_m\sigma_2 \bar\psi^\alpha_{Lb}\epsilon_{abe})
	(\psi^\beta_{Lc} \sigma_2\sigma_m \psi^\beta_{Ld}\epsilon_{cde})
	\notag
	\\
	& \quad + (L\leftrightarrow R)
	\Big]\,.
\ea
To bilinearize the quartic interaction we insert the constant factor
\ba
	&\int \rmd \Delta_L \exp\bigg[
		-8N \ckakko{(\Delta_L)_{me}-\frac{1}{8N}\bar\psi^\alpha_{La}\sigma_m\sigma_2 \bar\psi^\alpha_{Lb}\epsilon_{abe}}
	\notag
	\\
	&\times \ckakko{(\Delta_L^*)_{me}-\frac{1}{8N}\psi^\beta_{Lc}\sigma_2\sigma_m \psi^\beta_{Ld}\epsilon_{cde}}
	\bigg] \times (L\leftrightarrow R)
\ea
where $\Delta_{L,R}$ are complex $3\times 3$ matrices that transform as triplet under color $\SU(3)$ and spin $\SO(3)$. They are also charged under $\U(1)_B$ and $\U(1)_A$. Then
\begin{widetext}
\ba
	Z & \propto \int \rmd\Delta_L \int \rmd\Delta_R \int \rmd \bar\psi_R \rmd \bar\psi_L \rmd \psi_R \rmd \psi_L 
	\exp\big[
		-8N\Tr(\Delta_L^\dagger\Delta_L) - 8N\Tr(\Delta_R^\dagger\Delta_R)
		+ m\bar\psi_{Rai}^\alpha\psi_{Lai}^\alpha+m^*\bar\psi_{Lai}^\alpha\psi_{Rai}^\alpha
	\notag
	\\
	& \quad 
	+ \psi^\alpha_{La}\sigma_2\sigma_m \psi^\alpha_{Lb}\epsilon_{abc}\Delta_{Lmc}
	+ \bar\psi^\alpha_{La}\sigma_m\sigma_2 \bar\psi^\alpha_{Lb}\epsilon_{abc}\Delta^*_{Lmc}
	+ (L\leftrightarrow R)
	\big]
	\\
	& = \int \rmd\Delta_L \int \rmd\Delta_R 
	\exp[-8N\Tr(\Delta_L^\dagger\Delta_L) - 8N\Tr(\Delta_R^\dagger\Delta_R)]
	\int \rmd \bar\psi_R \rmd \bar\psi_L \rmd \psi_R \rmd \psi_L 
	\notag
	\\
	& \quad \times \exp\kkakko{
	\begin{pmatrix}\bar\psi_L \\ \bar\psi_R \\ \psi_L \\ \psi_R\end{pmatrix}^\alpha_{ai}
	\begin{pmatrix}
	(\sigma_m\sigma_2)_{ij}\epsilon_{abc}\Delta^*_{Lmc}&0&0&m^*\delta_{ab}\delta_{ij}/2
	\\
	0&(\sigma_m\sigma_2)_{ij}\epsilon_{abc}\Delta^*_{Rmc}&m\delta_{ab}\delta_{ij}/2&0
	\\
	0&-m\delta_{ab}\delta_{ij}/2&(\sigma_2\sigma_m)_{ij}\epsilon_{abc}\Delta_{Lmc}&0
	\\
	-m^*\delta_{ab}\delta_{ij}/2&0&0&(\sigma_2\sigma_m)_{ij}\epsilon_{abc}\Delta_{Rmc}
	\end{pmatrix}
	\begin{pmatrix}\bar\psi_L \\ \bar\psi_R \\ \psi_L \\ \psi_R\end{pmatrix}^\alpha_{bj}
	}
	\notag
	\\
	& = \int \rmd\Delta_L \int \rmd\Delta_R 
	\exp[-8N\Tr(\Delta_L^\dagger\Delta_L) - 8N\Tr(\Delta_R^\dagger\Delta_R)]
	\notag
	\\
	& \quad \times \Pf^N \begin{pmatrix}
	(\sigma_m\sigma_2)_{ij}\epsilon_{abc}\Delta^*_{Rmc}&m\delta_{ab}\delta_{ij}/2
	\\
	-m\delta_{ab}\delta_{ij}/2&(\sigma_2\sigma_m)_{ij}\epsilon_{abc}\Delta_{Lmc}
	\end{pmatrix}
	\Pf^N \begin{pmatrix}
	(\sigma_m\sigma_2)_{ij}\epsilon_{abc}\Delta^*_{Lmc} & m^*\delta_{ab}\delta_{ij}/2
	\\
	-m^*\delta_{ab}\delta_{ij}/2 & (\sigma_2\sigma_m)_{ij}\epsilon_{abc}\Delta_{Rmc}
	\end{pmatrix}.
\ea
\end{widetext}
This is an exact rewriting of the original partition function and so far no approximation has been made. Now we shall let $N$ large with $m\sim 1/\sqrt{N}$. To nail down the saddle point of the integral, let us take the chiral limit using the Pfaffian formulas \footnote{To verify these formulas we used SymPy, a Python library for symbolic mathematics \cite{10.7717/peerj-cs.103}.}
\ba
	\Pf[(\sigma_2\sigma_m)_{ij}\epsilon_{abc}\Delta_{mc}] & = 2 \det \Delta \,,
	\\
	\Pf[(\sigma_m\sigma_2)_{ij}\epsilon_{abc}\Delta_{mc}^*] & = 2 \det \Delta^\dagger  \,.
\ea
Here, $(\sigma_2\sigma_m)_{ij}\epsilon_{abc}\Delta_{mc}$ and $(\sigma_m\sigma_2)_{ij}\epsilon_{abc}\Delta_{mc}^*$ are treated as $6\times6$ matrices with left index $(a,i)$ and right index $(b,j)$. The partition function in the chiral limit therefore reads
\ba
	Z & \propto \int \rmd\Delta_L \int \rmd\Delta_R 
	\exp[-8N\Tr(\Delta_L^\dagger\Delta_L) - 8N\Tr(\Delta_R^\dagger\Delta_R)]
	\notag
	\\
	& \quad \times 
	{\det}^N(\Delta_L^\dagger \Delta_L){\det}^N(\Delta_R^\dagger \Delta_R)\,.
\ea
With a quick calculation involving the singular value decomposition of $\Delta_{L,R}$ one can verify that 
\ba
	\Delta_L = \Delta_R = \frac{1}{2\sqrt{2}}\1_3
	\label{eq:sadd}
\ea
maximizes the integrand. It is not invariant under separate rotations by color $\SU(3)$ and spin $\SO(3)$ but is invariant under their simultaneous rotation. Namely color and spin are locked:
\ba
	\U(1)_B\times \U(1)_A \times \SU(3)_C\times\SO(3)_J \to \SO(3)_{C+J}\,.
\ea
The soft fluctuations around \eqref{eq:sadd} can be parametrized as
\ba
	\Delta_L = U\,, \quad \Delta_R = V
\ea
with $U,V\in\U(3)$. (This $V$ should not be confused with the Gaussian random matrix $V^A$ inside $\mathscr{D}$.) Let us define
\ba
	\begin{split}
	\rme^{i\phi_U}\equiv\det U, \quad 
	\rme^{i\phi_V}\equiv\det V, 
	\\
	\mathcal{Q}_{ai,bj}\equiv\epsilon_{abm}(\sigma_2\sigma_m)_{ij}. \quad
	\end{split}
\ea
The large-$N$ partition function can be evaluated as follows:
\begin{widetext}
\ba
	Z &\sim 
	\int_{\U(3)}\!\!\!\!\rmd U \int_{\U(3)}\!\!\!\!\rmd V ~
	{\det}^{N/2} \begin{pmatrix}
	(\sigma_m\sigma_2)_{ij}\epsilon_{abc}V^*_{mc} & \sqrt{2}m\delta_{ab}\delta_{ij}
	\\
	-\sqrt{2}m\delta_{ab}\delta_{ij}&(\sigma_2\sigma_m)_{ij}\epsilon_{abc}U_{mc}
	\end{pmatrix}
	{\det}^{N/2} \begin{pmatrix}
	(\sigma_m\sigma_2)_{ij}\epsilon_{abc}U^*_{mc} & \sqrt{2} m^*\delta_{ab}\delta_{ij}
	\\
	- \sqrt{2}m^*\delta_{ab}\delta_{ij} & (\sigma_2\sigma_m)_{ij}\epsilon_{abc}V_{mc}
	\end{pmatrix}
	\\
	& = \int_{\U(3)}\!\!\!\!\rmd U \int_{\U(3)}\!\!\!\!\rmd V ~
	{\det}^{N/2} \kkakko{\begin{pmatrix}V^*\otimes\1_2&0\\0&U\otimes\1_2\end{pmatrix}
	\begin{pmatrix}
	\epsilon_{abc}V^*_{mc}(\sigma_m\sigma_2)_{ij} & \sqrt{2}m\delta_{ab}\delta_{ij}
	\\
	-\sqrt{2}m\delta_{ab}\delta_{ij} & \epsilon_{abc}U_{mc}(\sigma_2\sigma_m)_{ij}
	\end{pmatrix}
	\begin{pmatrix}V^\dagger\otimes\1_2&0\\0&U^T\otimes\1_2\end{pmatrix}}
	\notag
	\\
	& \quad \times 
	{\det}^{N/2} \kkakko{\begin{pmatrix}U^*\otimes\1_2&0\\0&V\otimes\1_2\end{pmatrix}
	\begin{pmatrix}
	\epsilon_{abc}U^*_{mc}(\sigma_m\sigma_2)_{ij} & \sqrt{2} m^*\delta_{ab}\delta_{ij}
	\\
	- \sqrt{2}m^*\delta_{ab}\delta_{ij} & \epsilon_{abc}V_{mc}(\sigma_2\sigma_m)_{ij}
	\end{pmatrix}
	\begin{pmatrix}U^\dagger\otimes\1_2&0\\0&V^T\otimes\1_2\end{pmatrix}}
	\\
	& \propto \int_{\U(3)}\!\!\!\!\rmd U \int_{\U(3)}\!\!\!\!\rmd V ~
	{\det}^{N/2}
	\begin{pmatrix}
	\rme^{-i\phi_V} \mathcal{Q}^* & \sqrt{2}m (V^*U^T)\otimes\1_2
	\\
	- \sqrt{2}m (UV^\dagger)\otimes\1_2 & \rme^{i\phi_U} \mathcal{Q}
	\end{pmatrix}
	\notag
	\\
	& \quad \times 
	{\det}^{N/2} 
	\begin{pmatrix}
	\rme^{-i\phi_U} \mathcal{Q}^* & \sqrt{2} m^*(U^*V^T)\otimes\1_2
	\\
	- \sqrt{2}m^* (VU^\dagger)\otimes\1_2 & \rme^{i\phi_V} \mathcal{Q}
	\end{pmatrix}
	\\
	& \propto \int_{\U(3)}\!\!\!\!\rmd U \int_{\U(3)}\!\!\!\!\rmd V ~
	{\det}^{N/2} 
	\begin{pmatrix}
	\1_6 & \sqrt{2}m [(V^*U^T)\otimes\1_2]\rme^{-i\phi_U} \mathcal{Q}^{-1}
	\\
	- \sqrt{2}m [(UV^\dagger)\otimes\1_2]\rme^{i\phi_V} \mathcal{Q}^{*-1} & \1_6
	\end{pmatrix}
	\notag
	\\
	& \quad \times 
	{\det}^{N/2} 
	\begin{pmatrix}
	\1_6 & \sqrt{2} m^*[(U^*V^T)\otimes\1_2]\rme^{-i\phi_V} \mathcal{Q}^{-1}
	\\
	- \sqrt{2}m^* [(VU^\dagger)\otimes\1_2]\rme^{i\phi_U} \mathcal{Q}^{*-1} & \1_6
	\end{pmatrix}
	\\
	& = \int_{\U(3)}\!\!\!\!\rmd U \int_{\U(3)}\!\!\!\!\rmd V ~
	{\det}^{N/2}\mkakko{
		\1_6 + 2m^2 \rme^{-i(\phi_U-\phi_V)}[(V^*U^T)\otimes\1_2]\mathcal{Q}^{-1}
		[(UV^\dagger)\otimes\1_2]\mathcal{Q}^{*-1}
	}
	\notag
	\\
	& \quad \times 
	{\det}^{N/2}\mkakko{
		\1_6 + 2m^{*2}\rme^{i(\phi_U-\phi_V)}[(U^*V^T)\otimes\1_2]\mathcal{Q}^{-1}[(VU^\dagger)\otimes\1_2]\mathcal{Q}^{*-1}
	}
	\\
	& \sim \int_{\U(3)}\!\!\!\!\rmd \hat{U}~
	\exp\kkakko{
		\hat{m}^2 \Tr \mkakko{[\hat{U}\otimes\1_2]\mathcal{Q}^{-1}
		[\hat{U}^T\otimes\1_2]\mathcal{Q}^{*-1}}
		+ \hat{m}^{*2}\Tr \mkakko{[\hat{U}^\dagger\otimes\1_2]\mathcal{Q}^{-1}
		[\hat{U}^*\otimes\1_2]\mathcal{Q}^{*-1}}
	}
	\label{eq:zzzsdf}
\ea
\end{widetext}
where we have defined 
\ba
	\begin{split}
	\hat{U}& \equiv \rme^{-i(\phi_U-\phi_V)/2}V^*U^T\,,
	\\
	\hat{m}& \equiv \sqrt{N}m\,.
	\end{split}
\ea
Using a software \cite{10.7717/peerj-cs.103} we verified
\ba
	& \Tr \mkakko{[\hat{U}\otimes\1_2]\mathcal{Q}^{-1}
	[\hat{U}^T\otimes\1_2]\mathcal{Q}^{*-1}} 
	\notag
	\\
	= & \frac{1}{2}\ckakko{
		\Tr(\hat{U}^2) - (\Tr \hat{U})^2 - \Tr (\hat{U}^T\hat{U})
	}.
	\label{eq:23243213}
\ea
Upon substituting \eqref{eq:23243213} into \eqref{eq:zzzsdf}, we finally arrive at the low-energy effective theory
\ba
	Z & \sim \int_{\U(3)}\!\!\!\!\rmd \hat{U} 
	\exp\bigg[
		\frac{\hat{m}^2}{2}\ckakko{
			\Tr(\hat{U}^2) - (\Tr \hat{U})^2 - \Tr (\hat{U}^T\hat{U})
		}
	\notag
	\\
	& \quad +\text{c.c.}
	\bigg]\,.
	\label{eq:3Zceff}
\ea
Notice that \eqref{eq:3Zceff} respects invariance under $\SO(3)$: $\hat{U}$ transforms as $\hat{U}\to g\hat{U}g^T$ for $g\in\SO(3)$. The $\U(1)$ part of $\hat{U}$ is the Nambu-Goldstone boson associated with the $\U(1)_A$ symmetry \footnote{Axial anomaly is suppressed at high density \cite{Schafer:2002ty}.}. On the other hand, the $\SU(3)$ part of $\hat{U}$ describes the Nambu-Goldstone bosons associated with the symmetry breaking $\SU(3)_C\times\SO(3)_J\to\SO(3)_{C+J}$. At this point the qualitative difference between QCD and RMT becomes evident. In one-flavor QCD, these 8 modes are absorbed into gluons via the Anderson-Higgs mechanism and disappear from the physical spectrum \cite{Schafer:2000tw,Pang:2010wk}. By contrast, there is no \emph{local} gauge invariance in RMT and the Anderson-Higgs mechanism is not operative, leading consequently to the appearance of these 8 modes in the low-energy effective theory \eqref{eq:3Zceff}. 

We also wish to point out that while our RMT concerns pairing of quarks with the same chirality, it is known that in one-flavor QCD pairing of quarks of opposite chirality can take place \cite{Pisarski:1999tv,Schafer:2000tw,Schmitt:2004et}. However, this is beyond the scope of this paper.

\section{\label{sc:Nc2}Two colors}

Due to its unique features and theoretical simplicity, two-color QCD has long been studied as a prototype of strongly coupled non-Abelian gauge theories. A particularly pleasant feature of two-color QCD is that for an even number of flavors with degenerate masses, the path-integral measure is positive definite even at nonzero quark chemical potential and can be simulated on a lattice \cite{Hands:1999md,Hands:2000ei,Kogut:2001na,Kogut:2002cm}. Interestingly, the color-singlet scalar diquarks in two-color QCD are degenerate with pions and condense when the chemical potential exceeds $m_{\pi}/2$, breaking $\U(1)_B$ spontaneously. The phase structure of two-color QCD has been studied with chiral perturbation theory \cite{Kogut:1999iv,Kogut:2000ek,Splittorff:2000mm,Splittorff:2001fy,Metlitski:2005db,Brauner:2006dv,Kanazawa:2009ks,Kanazawa:2011tt,Adhikari:2018kzh} and low-energy effective models \cite{Vanderheyden:2001gx,Klein:2004hv,Ratti:2004ra,Brauner:2009gu,Andersen:2010vu,He:2010nb,Strodthoff:2011tz,Imai:2012hr,Strodthoff:2013cua}. 

It seems that almost all previous work focused on two-color QCD with an even number of flavors \footnote{In \cite{Hands:2000ei} two-color QCD with a single flavor in the \emph{adjoint} representation of $\SU(2)$ was investigated.}, but there is a qualitative difference between even and odd flavors. (Note that we refer to the number of \emph{Dirac} fermions here. The famous Witten's $\SU(2)$ anomaly \cite{Witten:1982fp}, which asserts that an $\SU(2)$ gauge theory coupled to an odd number of \emph{Weyl} fermions in the spin-$1/2$ representation of $\SU(2)$ is inconsistent, does not affect our present discussion.) It is easy to understand this intuitively. If we assume that quarks form pairs in the Lorentz-scalar color-antisymmetric channel, the Pauli principle stipulates that the quantum number of flavors must be antisymmetric. For even flavors all quarks can participate in this pairing, while for odd flavors a single flavor remains gapless (recall that an antisymmetric matrix of odd dimension always has a zero eigenvalue). Then what option is available for this isolated quark? One option is to form a color-adjoint diquark condensate in the Lorentz-scalar channel, causing color superconductivity. At least in the high-density limit we regard this possibility as unlikely, because the single-gluon exchange interaction is repulsive in the color-symmetric channel. Another option is to form a color-singlet diquark condensate in the channel of total angular momentum $J=1$. It inevitably breaks rotational symmetry. Neither a low-energy effective theory nor RMT is known for this phase. 

In order to understand implications of our RMT for dense two-color QCD, let us return to \eqref{eq:Z87}. For $N_c=2$ the identity $\epsilon_{ab}\epsilon_{cd} = \delta_{ac}\delta_{bd}-\delta_{ad}\delta_{bc}$ holds. Therefore
\ba
	Z & \propto 
	\int \rmd \bar\psi_R \rmd \bar\psi_L \rmd \psi_R \rmd \psi_L 
	\exp\Big[
		m\bar\psi_{Rai}^\alpha\psi_{Lai}^\alpha+m^*\bar\psi_{Lai}^\alpha\psi_{Rai}^\alpha
	\notag
	\\
	& \quad 
	+ \frac{1}{8N}(\bar\psi^\alpha_{La} \sigma_m\sigma_2 \epsilon_{ab} \bar\psi^\alpha_{Lb})
	(\psi^\beta_{Lc} \sigma_2\sigma_m \epsilon_{cd} \psi^\beta_{Ld}) 
	\notag
	\\
	& \quad + (L\leftrightarrow R) \Big]\,.
\ea
The ensuing analysis is routine. To bilinearize the quartic interaction we insert the constant factor
\ba
	& \int_{\CC^3} \rmd \vec\Omega_L 
	\exp\bigg[
		-8N\mkakko{\Omega_{Lm}-\frac{1}{8N}\bar\psi^\alpha_{La}\sigma_m\sigma_2 \epsilon_{ab} \bar\psi^\alpha_{Lb}}
	\notag
	\\
	& \times 
		\mkakko{\Omega^*_{Lm}-\frac{1}{8N}\psi^\beta_{Lc}\sigma_2\sigma_m \epsilon_{cd} \psi^\beta_{Ld}}
	\bigg] \times(L\leftrightarrow R) 
\ea
where $\vec\Omega_{L,R}$ are three-component complex vectors that transform as triplet under $\SO(3)$ and 
are also charged under $\U(1)_B\times\U(1)_A$. Then
\begin{widetext}
\ba
	Z & \propto 
	\int_{\CC^3} \rmd \vec\Omega_L \int_{\CC^3} \rmd \vec\Omega_R 
	\int \rmd \bar\psi_R \rmd \bar\psi_L \rmd \psi_R \rmd \psi_L ~
	\exp\bigg[
		-8N(|\vec\Omega_L|^2+|\vec\Omega_R|^2) + 
		m\bar\psi_{Rai}^\alpha\psi_{Lai}^\alpha+m^*\bar\psi_{Lai}^\alpha\psi_{Rai}^\alpha
	\notag
	\\
	& \quad 
	+ \Omega_{Lm}\psi^\alpha_{La}\sigma_2\sigma_m \epsilon_{ab} \psi^\alpha_{Lb}
	+ \Omega^*_{Lm}\bar\psi^\alpha_{La}\sigma_m\sigma_2 \epsilon_{ab}\bar\psi^\alpha_{Lb}
	+ (L\leftrightarrow R)
	\bigg]
	\\
	& = \int_{\CC^3} \rmd \vec\Omega_L \int_{\CC^3} \rmd \vec\Omega_R 
	\int \rmd \bar\psi_R \rmd \bar\psi_L \rmd \psi_R \rmd \psi_L ~
	\exp[-8N(|\vec\Omega_L|^2+|\vec\Omega_R|^2)]
	\notag
	\\
	& \quad \times \exp\kkakko{
		\begin{pmatrix}\bar\psi_L\\\bar\psi_R\\\psi_L\\\psi_R\end{pmatrix}^\alpha_{ai}
		\begin{pmatrix}
			\Omega^*_{Lm}(\sigma_m\sigma_2)_{ij}\epsilon_{ab}&0&0&m^*\delta_{ab}\delta_{ij}/2
			\\
			0&\Omega^*_{Rm}(\sigma_m\sigma_2)_{ij}\epsilon_{ab}&m\delta_{ab}\delta_{ij}/2&0
			\\
			0&-m\delta_{ab}\delta_{ij}/2&\Omega_{Lm}(\sigma_2\sigma_m)_{ij}\epsilon_{ab}&0
			\\
			-m^*\delta_{ab}\delta_{ij}/2&0&0&\Omega_{Rm}(\sigma_2\sigma_m)_{ij}\epsilon_{ab}
		\end{pmatrix}
		\begin{pmatrix}\bar\psi_L\\\bar\psi_R\\\psi_L\\\psi_R\end{pmatrix}^\alpha_{bj}
	}
	\\
	& = \int_{\CC^3} \rmd \vec\Omega_L \int_{\CC^3} \rmd \vec\Omega_R ~
	\exp[-8N(|\vec\Omega_L|^2+|\vec\Omega_R|^2)]
	\notag
	\\
	& \quad \times \Pf^N\begin{pmatrix}
	\Omega^*_{Rm}(\sigma_m\sigma_2)_{ij}\epsilon_{ab}&m\delta_{ab}\delta_{ij}/2
	\\
	-m\delta_{ab}\delta_{ij}/2&\Omega_{Lm}(\sigma_2\sigma_m)_{ij}\epsilon_{ab}
	\end{pmatrix}
	\Pf^N\begin{pmatrix}
	\Omega^*_{Lm}(\sigma_m\sigma_2)_{ij}\epsilon_{ab} & m^*\delta_{ab}\delta_{ij}/2
	\\
	-m^*\delta_{ab}\delta_{ij}/2 & \Omega_{Rm}(\sigma_2\sigma_m)_{ij}\epsilon_{ab}
	\end{pmatrix}
	\\
	& = \int_{\CC^3} \rmd \vec\Omega_L \int_{\CC^3} \rmd \vec\Omega_R ~
	\exp[-8N(|\vec\Omega_L|^2+|\vec\Omega_R|^2)]
	\notag
	\\
	& \quad \times 
	{\det}^N \begin{pmatrix}
	\Omega^*_{Rm}(\sigma_m\sigma_2)_{ij}&m\delta_{ij}/2
	\\
	m\delta_{ij}/2&\Omega_{Lm}(\sigma_2\sigma_m)_{ij}
	\end{pmatrix}
	{\det}^N \begin{pmatrix}
	\Omega^*_{Lm}(\sigma_m\sigma_2)_{ij}&m^*\delta_{ij}/2
	\\
	m^*\delta_{ij}/2&\Omega_{Rm}(\sigma_2\sigma_m)_{ij}
	\end{pmatrix}
	\\
	& = \int_{\CC^3} \rmd \vec\Omega_L \int_{\CC^3} \rmd \vec\Omega_R ~
	\exp[-8N(|\vec\Omega_L|^2+|\vec\Omega_R|^2)]
	\ckakko{\vec\Omega_R^{*2} \vec\Omega_L^2 
	- \frac{m^2}{2}\vec\Omega_R^*\cdot\vec\Omega_L+\frac{m^4}{16}}^N
	\ckakko{\vec\Omega_L^{*2}\vec\Omega_R^2 
	- \frac{m^{*2}}{2}\vec\Omega_L^*\cdot\vec\Omega_R+\frac{m^{*4}}{16}}^N.
	\label{eq:z024875}
\ea
\end{widetext}
We are now ready to take the microscopic large-$N$ limit with $m\propto 1/\sqrt{N}$. In the chiral limit we find
\ba
	Z & \propto \int_{\CC^3} \rmd \vec\Omega_L \int_{\CC^3} \rmd \vec\Omega_R ~
	\exp[-8N(|\vec\Omega_L|^2+|\vec\Omega_R|^2)]
	\notag
	\\
	& \quad \times |\vec\Omega_L^2|^{2N}|\vec\Omega_R^2|^{2N}
	\\
	& = \ckakko{\int_{\CC^3}\rmd \vec\Omega ~
	\big(\rme^{-8|\vec\Omega|^2}|\vec\Omega^2|^2\big)^N}^2.
\ea
Thus the problem of finding the saddle point at large $N$ boils down to extremizing the function $f(\vec\Omega)\equiv\rme^{-8|\vec\Omega|^2}|\vec\Omega^2|^2$. To simplify the problem we shall draw upon the fact that by a suitable $\U(1)\times\SO(3)$ rotation, an arbitrary complex 3-vector $\vec\Omega$ can be brought to the form $(a+ib,c,0)$ with $a,b,c\in\RR$. Then
\ba
	f(\vec\Omega) & = \rme^{-8(a^2+b^2+c^2)}\kkakko{(a^2-b^2+c^2)^2+4a^2b^2}.
\ea
This function is maximized when $b=0$ and $a^2+c^2=1/4$, or when $c=0$ and $a^2+b^2=1/4$. To show this, first note that $f = \rme^{-8(a^2+b^2+c^2)}\kkakko{(a^2+b^2+c^2)^2-4b^2c^2}$. This means that $f$ can be increased by rotating $(b,c)$ into either $(\sqrt{b^2+c^2},0)$ or $(0,\sqrt{b^2+c^2})$. Thus one can safely assume $bc=0$. If $b=0$, the vector $(a,c,0)$ can be rotated to $(\sqrt{a^2+c^2},0,0)$ by the action of $\SO(3)$. If $c=0$, the vector $(a+ib,0,0)$ can be rotated to $(\sqrt{a^2+b^2},0,0)$ by the action of $\U(1)$. In either case, a quick calculation shows that $(1/2,0,0)$ maximizes $f$.

The above analysis reveals that the manifold of maxima of $f$ is given by $\U(1)\times\SO(3)$ rotations of $(1/2,0,0)$. Picking out any single point on this manifold as a ground state breaks global symmetries according to
\ba
	\U(1)\times\SO(3) \to \SO(2).
\ea
Note that this breaking occurs for each chirality independently. 

The low-energy fluctuations in the ground state can be parametrized as
\ba
	\vec\Omega_L = \frac{1}{2}\rme^{i\phi_L}\n_L
	\quad \text{and}\quad 
	\vec\Omega_R = \frac{1}{2}\rme^{i\phi_R}\n_R
\ea
where the unit vectors $\n_{L,R}\in\RR^3$ $(\n_{L,R}^2=1)$ stand for the directions of diquarks in the spin space. For $N\gg 1$ with $\hat{m}\equiv \sqrt{N}m \sim O(1)$, we have from \eqref{eq:z024875}
\ba
	Z & \sim \int_{\U(1)}\!\!\!\! \rmd \phi_L \int_{\U(1)} \!\!\!\! \rmd \phi_R \int_{S^2}\rmd \n_L \int_{S^2}\rmd \n_R 
	\notag
	\\
	& \quad \times 
	\ckakko{\frac{1}{16}\rme^{2i(\phi_L-\phi_R)}-\frac{m^2}{8}\rme^{i(\phi_L-\phi_R)}\n_L\cdot \n_R}^N
	\notag
	\\
	& \quad \times 
	\ckakko{\frac{1}{16}\rme^{-2i(\phi_L-\phi_R)}-\frac{m^{*2}}{8}\rme^{-i(\phi_L-\phi_R)}\n_L\cdot \n_R}^N
	\notag
	\\
	& \sim \int_{\U(1)}\!\!\!\! \rmd \phi_L \int_{\U(1)} \!\!\!\! \rmd \phi_R \int_{S^2}\rmd \n_L \int_{S^2}\rmd \n_R 
	\notag
	\\
	& \quad \times
	\exp\big[
		-2\hat{m}^2 \rme^{-i(\phi_L-\phi_R)}\n_L\cdot \n_R + \text{c.c.}
	\big]
	\\
	& \propto 
	\int_{\U(1)}\!\!\!\! \rmd \phi \int_{S^2}\rmd \n_L \int_{S^2}\rmd \n_R ~
	\exp\mkakko{ 4|\hat{m}|^2 \n_L\cdot \n_R \cos\phi }.
\ea
The $\phi$ field is nothing but the Nambu-Goldstone mode for the $\U(1)_A$ symmetry. Integrating out $\phi$ yields
\ba
	Z & \propto \int_{S^2}\rmd \n_L \int_{S^2}\rmd \n_R~ I_0(4|\hat{m}|^2 \n_L\cdot \n_R)
	\label{eq:025840875}
	\\
	& \propto \int_0^\pi \rmd \theta \sin\theta~I_0(4|\hat{m}|^2 \cos\theta)
	\\
	& \propto \int_{0}^{1}\rmd x~I_0(4x|\hat{m}|^2)\,,
\ea
where $I_0$ is the modified Bessel function of the first kind. This is the simplest form we found for the microscopic partition function. An important observation is that a nonzero quark mass tends to make the diquarks of opposite chiralities parallel to each other. In fact, for a large mass $|\m|\gg 1$, the leading asymptotic behavior $I_0(z)\sim \rme^{|z|}/\sqrt{2\pi |z|}$ for $|z|\gg1$ \cite{NIST:DLMF}, as applied to \eqref{eq:025840875}, shows that a dominant contribution to $Z$ comes from the region with $\n_L \cdot \n_R\sim \pm 1$, i.e., $\n_L\parallel \n_R$.

\section{\label{sc:conc}Conclusions}

In this paper we reported an attempt to extend chiral random matrix theory (RMT) \cite{Verbaarschot:2000dy} to cold and dense one-flavor quark matter, in which quarks are thought to form Cooper pairs with nonzero total angular momentum \cite{Bailin:1983bm}. For three colors the putative ground state is the color-spin locked (CSL) phase \cite{Schafer:2000tw} that breaks color and rotational symmetries but retains invariance under their combination, bearing a close resemblance to the B phase of $^3$He and the CFL phase of three-flavor quark matter \cite{Alford:1998mk}. In this paper we have formulated a new non-Hermitian chiral RMT and showed that in the microscopic large-$N$ limit it exhibits the same symmetry breaking pattern as the CSL phase of QCD. This work can be viewed as a natural sequel to our previous work \cite{Kanazawa:2020dju} that put forward chiral RMT for Cooper pairing in two and three-flavor dense QCD. A downside of our RMT is the existence of 8 unphysical Nambu-Goldstone modes that enter into the large-$N$ low-energy effective theory. They are associated with the breaking of color $\SU(3)$ symmetry. Although they are ``eaten'' by gluons in the CSL phase of QCD, they do represent physical fluctuations in RMT because the Anderson-Higgs mechanism does not operate in a zero-dimensional theory with no local gauge invariance.  
In the second half of this paper we also discussed the application of our RMT to two colors. In this case the symmetry breaking pattern of RMT was found to be $\U(1)\times\SO(3)\to\SO(2)$, which matches physical expectations for two-color one-flavor QCD at high density. It would be worthwhile to perform weak-coupling calculations of gap equations for the latter in order to precisely check the validity of correspondence between our RMT and two-color QCD. Analytical calculation of the microscopic spectral density of the Dirac operator $\mathscr{D}$ in RMT is quite challenging and we leave it to future work.

%\bibliographystyle{apsrev4-2}
%\bibliography{../physics_cflrmt/refs_rmt}
\bibliography{draft_CSL.bbl}
\end{document}